\documentclass[12pt,a4paper,dvips,authordate1-4]{article}
\usepackage{epsfig,wrapfig,times,mathptm} 
\renewcommand{\section}[1]{\vspace{6pt} \noindent\mbox{#1} \newline \noindent}
\renewcommand{\subsection}[1]{\vspace{6pt} \noindent\mbox{\underline{#1}} 
\newline \noindent}
\renewcommand{\subsubsection}[1]{\vspace{6pt} \noindent\mbox{\underline{#1}}
\noindent}

\newfont{\sansb}{cmssbx10}
\newfont{\sans}{cmss10}

\begin{document}
{\small OG 4.3.4 \vspace{-24pt}\\}     
{\center \LARGE TeV OBSERVATIONS OF THE VARIABILITY AND SPECTRUM OF MARKARIAN 421
\vspace{6pt}\\}

J.~E.~McEnery,$^1$
I.~H.~Bond,$^2$
P.~J.~Boyle,$^1$
S.~M.~Bradbury,$^2$
A.~C.~Breslin,$^1$
J.~H.~Buckley,$^3$
A.~M.~Burdett,$^2$
J.~Bussons~Gordo,$^1$
D.~A.~Carter-Lewis,$^4$
M.~Catanese,$^4$
M.~F.~Cawley,$^5$
D.~J.~Fegan,$^1$
J.~P.~Finley,$^6$
J.~Gaidos,$^6$
A.~Hall,$^6$
A.~M.~Hillas,$^2$
F.~Krennrich,$^4$
R.~C.~Lamb,$^7$
R.~W.~Lessard,$^6$
C.~Masterson,$^1$
G.~Mohanty,$^{4}$
P.~Moriarty,$^{8}$
J.~Quinn,$^{3,1}$
A.~J.~Rodgers,$^{2}$
H.~J.~Rose,$^{2}$
F.~W.~Samuelson,$^4$
G.~H.~Sembroski,$^{6}$
R.~Srinivasan,$^6$
T.~C.~Weekes,$^{3}$
and J.~Zweerink$^{4}$\\
$^1$ Physics Department, University College, Dublin
4, Ireland\\
$^2$ Department of Physics, University of Leeds,
Leeds, LS2 9JT, Yorkshire, England, UK\\
$^3$ Fred Lawrence Whipple Observatory, Harvard-
Smithsonian CfA, P.O. Box 97, Amado, AZ 85645\\ 
$^4$ Department of Physics and Astronomy, Iowa State
University, Ames, IA 50011-3160\\
$^5$ Physics Department, St.Patrick's College,
Maynooth, County Kildare, Ireland\\
$^6$ Department of Physics, Purdue University, West
Lafayette, IN 47907\\
$^7$ Space Radiation Lab, Caltech, Pasadena, CA 91125 \\
$^8$ Department of Physical Sciences, Regional Technical College, Galway, Ireland
\vspace{6pt}

{\center ABSTRACT\\} 
Markarian 421 was the first extragalactic source to be detected with
high statistical certainty at TeV energies. The Whipple Observatory
gamma-ray telescope has been used to observe the Active Galactic
Nucleus, Markarian 421 in 1996 and 1997. The rapid variability
observed in TeV gamma rays in previous years is confirmed.  Doubling
times as short as 15 minutes are reported with flux levels reaching 15
photons per minute. The TeV energy spectrum is derived using two
independent methods. The implications for the intergalactic infra-red
medium of an observed unbroken power law spectrum up to energies of 5
TeV is discussed.

\setlength{\parindent}{1cm}
\section{INTRODUCTION}
Markarian 421, a nearby (z=0.031), X-ray selected BL Lacertae object
was the first extragalactic object to be detected at energies greater
than 30 GeV (Punch et al, 1992). BL Lacertae objects are a subset of a
broader class of Active Galactic Nuclei (AGN) known as blazars. It is
widely believed that the characteristic properties of blazars (rapid
variability over a wide range of energies, flat radio spectrum and
high and variable polarization) result because the jets in these AGN
are closely aligned to the line of sight (Blandford and K\"{o}nigl,
1979). The rapid $\gamma$-ray variability reported in this paper
supports this interpretation since relativistic beaming is necessary
to avoid absorption of the highest energy radiation.

The TeV $\gamma$-ray emission from Markarian 421 has been monitored
from December through May in 1996 and 1997.  The observations were
made with the 10 meter $\gamma$-ray telescope at the Whipple
Observatory. The camera was significantly upgraded during Autumn 1996
to 151 pixels increasing the field of view of the telescope from
3$^{\circ}$ to 3.5$^{\circ}$. The imaging atmospheric technique uses
information on the angular shape and orientation of individual
\v{C}erenkov images to reject more than 99.7\% of the cosmic-ray induced
showers while retaining over 50\% of the possible source $\gamma$-ray
initiated events (Reynolds et al, 1993). Since the instrument has
changed significantly between the 1996 and 1997 observing seasons, the
$\gamma$-ray flux for each season is quoted as a fraction of the
$\gamma$-ray flux from the Crab Nebula so that data from the two
seasons may be meaningfully compared.

\section{FLUX VARIABILITY: DECEMBER 1995-MAY 1997}
\begin{figure}[h!]
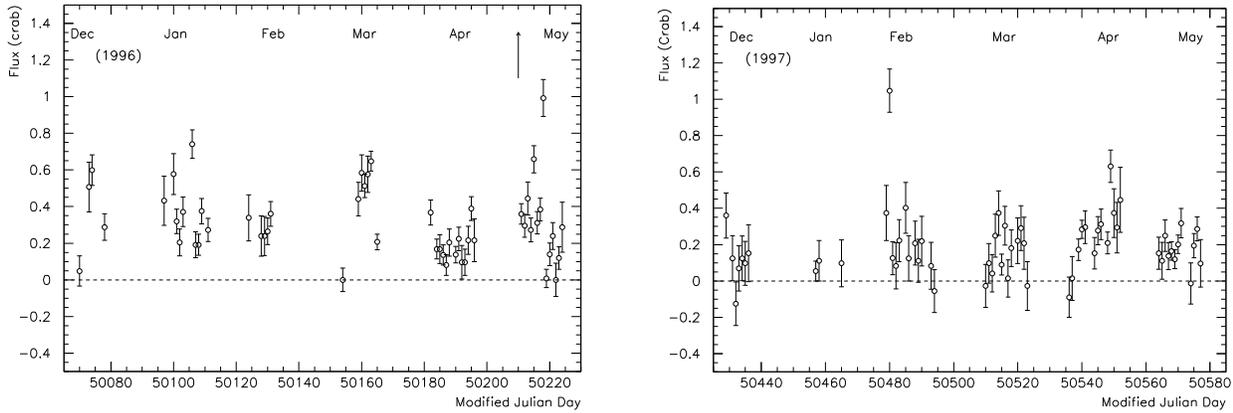

\parbox[t]{8.4cm}{
\centerline{\epsfig{figure=m4_icrc1.epsi,width=7.6cm}}}
\ \
\parbox[b]{8.4cm}{
\centerline{\epsfig{figure=m4_icrc2.epsi,width=7.6cm}}
}
\caption{The daily $\gamma$-ray rates during 1996 and 1997 as a fraction of 
the rate from the Crab Nebula, data from May 7, 1996 (6.2 Crab) is supressed
and is represented here by an arrow.}
\label{fig:rate_m4}
\end{figure}

The $\gamma$-ray emission averaged over each observing season has
gradually decreased since the 1995 observing season from 0.35 Crab in
1995 to 0.2 Crab in 1997. The $\gamma$-ray emission from Mrk421
exhibits a significant degree of dayscale flickering, the $\chi ^{2}$
probability that the daily $\gamma$-ray rate from Mrk421 in 1996 was
constant about the average rate is $3 \times 10^{-36}$. A similar test
applied to data from the Crab Nebula taken in 1996 gives a probability
of 0.86 for steady emission.  This result shows that the emission from
the Crab Nebula is constant and indicates that fluctuations due to
systematic effects are negligible compared to the statistical errors.
Figure~\ref{fig:rate_m4} shows the $\gamma$-ray flux from Mrk421 as a
fraction of the Crab Nebula flux for the 1996 and 1997 observing
seasons.


\begin{figure}[h!]
\centerline{\epsfig{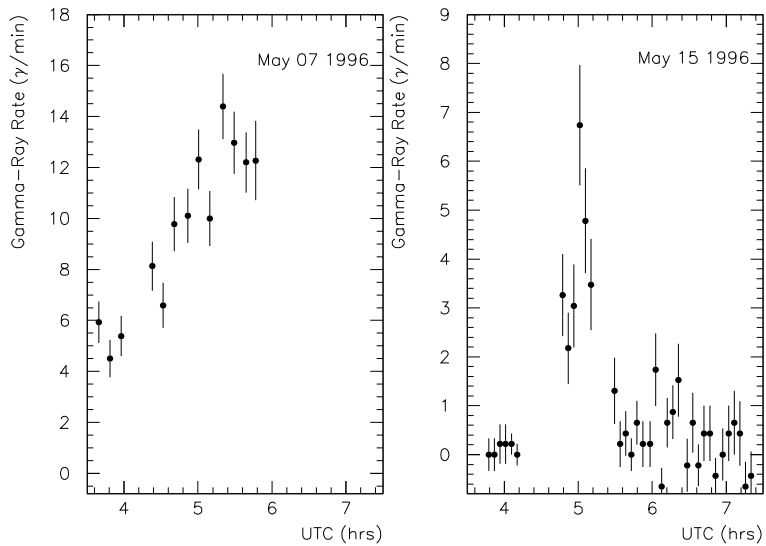}}
\caption{Light curves of two Mrk 421 flare events of 1996 May 7 and 1996 May 15.}
\label{fig:huge}
\end{figure}

A number of well defined flares can be seen in these data. On February
1, 1997, the $\gamma$-ray emission increased by a factor of 5.5 above
the average flux for the 1997 observing season.  No evidence for
hourscale variability within the 1 hour of observations on this night
was found. The most impressive variations observed occurred in May
1996 when, in addition to an increase of dayscale flickering, two
dramatic hourscale flares were observed.  These are shown in
Figure~\ref{fig:huge} (Gaidos et al, 1996). The first occurred on May
7, where the flux increased by a factor of 5 in the 2.5 hours of
observations and reached a maximum rate $\sim$10 times the Crab,
becoming the most intense source of TeV $\gamma$-rays ever observed.
The second, on May 15, was only $\sim$ 30 minutes in duration, with a
peak TeV flux of about 14 times the average for 1996.

\section{SPECTRUM}
\begin{figure}[h!]
\centerline{\epsfig{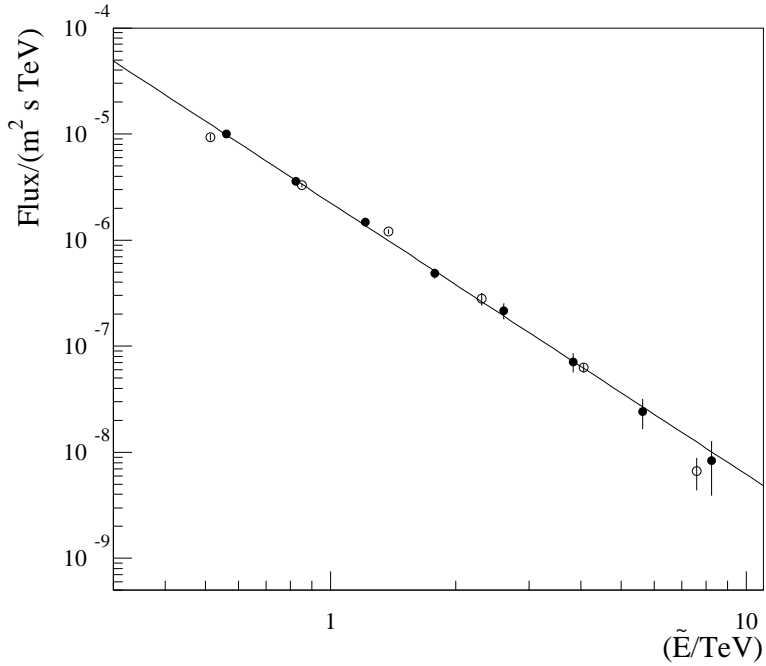}}
\caption{Differential energy spectrum of Markarian 421 fom observations taken
on May 7 1996. The filled circles are calculated using method 1
(Zweerink et al 1997) and the open circles using method 2 (Rodgers et
al, 1997)}
\label{fig:specm4}
\end{figure}

The extremely large $\gamma$-ray flux observed on May 7 from Mrk421
provides a very pure sample of $\gamma$-rays from which we can extract
an energy spectrum and search for evidence for a spectral cutoff. The
spectrum of Mrk421 was obtained using two independent analysis methods
described in detail as method 1 and method 2 of Mohanty et al (1997).
Method 1 isolates $\gamma$-rays from the dominant background of
cosmic-ray induced images using parameter cuts which have a slight
energy dependence. This compensates for the fact that showers from
higher energy $\gamma$-rays have slightly larger images.  Method 2
uses a ``cluster'' or ``spherical'' approach in which a single
parameter is used to characterize the $\gamma$-ray-like nature of an
image and correlations between image parameters are incorporated
naturally. Figure~\ref{fig:specm4} shows a plot of the $\gamma$-ray
flux versus energy using method 1 (filled circles, Zweerink et al. 1997)
and method 2 (open circles, Rodgers 1997) along with the best fit
power law of:
\begin{equation}
F(E) = (2.24\pm0.12\pm0.7) \times 10^{-6}E^{-2.56\pm 0.07\pm 0.1} {\rm photons/s/m^{2}/TeV.}
\end{equation}

In June 1995, a considerable amount of data was taken at large zenith angles
while Mrk421 was in a high $\gamma$-ray state. The large zenith angles
cause the energy region covered by the observations to shift to
significantly higher energies relative to the standard small zenith
angle observations. Monte Carlo simulations were used to determine the
energy threshold, collection area and optimal parameter cuts at a
variety of large zenith angles. Large zenith angle observations of the
Crab Nebula were used to perform a consistency check of the
predictions of the Monte Carlo simulation regarding the energy
threshold and collection areas at different zenith angles. (Krennrich
et al, 1997)

Observations at zenith angles of 45$^{\circ}$-50$^{\circ}$ on June 20, 21 
and 28 and 55$^{\circ}$-60$^{\circ}$ on June 20 were used to search for a 
spectral cut-off in the very high energy $\gamma$-ray emission from Mrk421.
Table~\ref{tab:lzen} shows the excess events at different energy
thresholds, it is evident that the spectrum of Mrk421 extends beyond
$5\pm 1.5$ TeV. Although the statistical significance above $8\pm 2.4$
TeV is marginal ($3.1\sigma$), a hint that the emission extends beyond
$8\pm 2.4$ TeV is present in the data.

\begin{table}[h]
\caption{Number of excess events from June 20,21,28 at large zenith angles (Krennrich et al, 1997)}
\label{tab:lzen}
\begin{center}
\begin{tabular}{lcc}
\hline\hline
Energy Threshold (TeV) & Excess Events & significance ($\sigma$) \\
\hline
$2.0\pm 0.6$ & 109 & 9.3 \\
$4.0\pm 1.2$ & 41  & 6.0 \\
$5.0\pm 1.5$ & 25  & 5.0 \\
$8.0\pm 2.4$ & 11  & 3.1 \\
\hline 
\hline
\end{tabular}
\end{center}
\end{table}

\section{DISCUSSION}
The extremely rapid $\gamma$-ray variability observed in May 1996
implies, by relativistic causality, a very compact emission region. A
significant degree of Doppler boosting is required to avoid absorption
of the $\gamma$-rays by pair production off low energy photon fields.
A lower limit of 9.9 can be derived for the Doppler boost factor of
the $\gamma$-ray emission region in the jet by considering the
simultaneous optical U-band flux (Buckley et al, 1996b).

Gamma-ray emission with E$\ge$ 5 TeV from Mrk421 has been detected using
two independent methods and datasets; no evidence for a cutoff in the
energy spectrum is seen in the data. This result conflicts with the
interpretation of a previously reported energy spectrum (Mohanty et
al, 1993) by De Jager, Stecker, and Salomon (1994), where they attributed a
non-detection at energies above 5 TeV to a cutoff by extragalactic
IR background radiation.

\section{ACKNOWLEDGEMENTS}
This research is supported by grants from the U.S. Department of Energy
and by NASA, by PPARC in the UK and by Forbairt in Ireland.


\section{REFERENCES}
\setlength{\parindent}{-5mm}
\begin{list}{}{\topsep 0pt \partopsep 0pt \itemsep 0pt \leftmargin 5mm
\parsep 0pt \itemindent -5mm}
\vspace{-15pt}
\item Blandford, R.D. and K\"{o}nigl, A., ApJ, 232, 34 (1979)
\item Buckley, J.H., et al, ApJ, 472, L9 (1996)
\item Buckley, J.H., et al, Adv. in Space Research, (1996b) (in press)
\item De Jager, O.C., Stecker, F.W. and Salamon, M.H., Nature, 369, 294 (1994)
\item Gaidos, J.A., et al, Nature, 383, 319 (1996)
\item Krennich, F.K. et al, ApJ, 481, 758 (1997)
\item Mohanty, G. et al, (in preparation) (1997)
\item Punch, M., et al, Nature, 358, 477 (1992)
\item Reynolds, P.T. et al., ApJ, 404, 206 (1993)
\item Rodgers, A.J., Private communication (1997)
\item Zweerink, J.A. et al, (in preparation) (1997)
\end{list}

\end{document}